\documentclass[aps, twocolumn, a4paper, superscriptaddress, 10pt, pre, floatfix]{revtex4-2}

\usepackage{graphicx}
\usepackage{amsmath, amssymb}
\usepackage{lmodern}
\usepackage[normalem]{ulem}
\usepackage[T1]{fontenc}
\usepackage[utf8]{inputenc}
\usepackage[english]{babel}
\usepackage{xcolor}

\begin{document}

\title{Monitoring biodiversity on highly reactive rock-paper-scissors models}

\author{D. Bazeia}
\affiliation{Departamento de Física, Universidade Federal da Paraíba, 58051-970 João Pessoa, PB, Brazil}
\author{M. J. B. Ferreira}
\affiliation{Departamento de Física, Universidade Estadual de Maringá, 87020-900 Maringá, PR, Brazil}
\author{B. F. de Oliveira}
\affiliation{Departamento de Física, Universidade Estadual de Maringá, 87020-900 Maringá, PR, Brazil}
\author{W. A. dos Santos}
\affiliation{Departamento de Física, Universidade Estadual de Maringá, 87020-900 Maringá, PR, Brazil}

\begin{abstract}
This work investigates how biodiversity is affected in a cyclic spatial May-Leonard model with hierarchical and non-hierarchical rules. Here we propose a generalization of the traditional rock-paper-scissors model by considering highly reactive species, i. e., species that react in a stronger manner compared to the others in respect to either competition or reproduction. These two classes of models, called here Highly Competitive and Highly Reproductive models, may lead to hierarchical and non-hierarchical dynamics, depending on the number of highly reactive species. The fundamental feature of these models is the fact that hierarchical models may as well support biodiversity, however, with a higher probability of extinction than the non-hierarchical ones, which are in fact more robust. This analysis is done by evaluating the probability of extinction as a function of mobility. In particular, we have analyzed how the dominance scheme changes depending on the highly reactive species for non-hierarchical models, where the findings lead to the conclusion that highly reactive species are usually at a disadvantage compared to the others. Moreover, we have investigated the power spectrum and the characteristic length of each species, including more information on the behavior of the several systems considered in the present work.
\end{abstract}

\maketitle

\section{Introduction}

It is quite common in nature examples of ecological systems formed by different species of living individuals coexisting and interacting among themselves; see, e.g., Refs. \cite{1992-Nowak-N-359-826, 1994-Jacob-N-368-46, 1998-Burrows-MEPS-167-1, 2002-Kerr-N-418-171,2004-Kirkup-N-428-412, 2006-Nowak-book, 2007-Reichenbach-N-488-1046,2020-Liao-NC-11-6065} and references therein. {\color{black} Despite the several achievements brought to light in the last decades, it remains an open question when it comes to the mechanisms that act behind biodiversity. In this field, great effort has been focused on providing a better understanding of the rules and mechanisms to control biodiversity }\cite{2014-Szolnoki-JRSI-11-100,2011-Jiang-PRE-84-021912,2015-Szolnoki-NJP-17-113033,2016-Szolnoki-SR-6-38608}. On the one hand, competition among individuals of the same species may occur, for example, when seeking for food, water, or shelter. On the other hand, competitions may also occur among individuals of different species, for example, in the presence of predator-prey interactions \cite{1987-Kuno-inbook}. Competition are also present in games of cooperators and defectors as in the public goods game \cite{2013-Szolnoki-PRX-3-041021,2012-Szolnoki-PRL-109-078701}. It may seen surprising at first glance, but competitions are central in the emergence and maintenance of biodiversity \cite{1920-Lokta-PNAS-6-410, 1926-Volterra-N-118-558, 1975-May-SIAMJAM-29-243, 1978-Wangersky-ARES-9-189}. For that reason, it is important to understand the fundamental mechanisms behind these competitions.

\begin{figure}[!htp]
    \centering
    \includegraphics{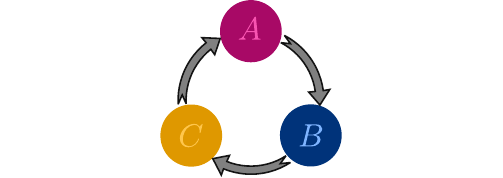}
    \caption{
        Non-hierarchical (or symmetric) competition rule among species. In this scheme, species $A$ dominates species $B$, species $B$ dominates species $C$, and species $C$ dominates species $A$, within a cyclic and symmetric environment.}
    \label{fig1}
\end{figure}

Due to the complexity of natural ecological systems, it would be a hard task to find a model that captures the whole essence of any real system. Thus, simpler models are sometimes implemented as attempts to describe the emergence of some basic properties and behaviors that systems supporting biodiversity can engender \cite{1991-Boerlijst-PD-48-17, 1991-Tainaka-EPL-15-399, 1994-Satulovsky-PRE-49-5073, 2007-Washenberger-JPCM-19-065139}. In this context, one important class of models engenders the so-called non-hierarchical models, of which the rock-paper-scissors (RPS) is a well-established example \cite{2002-Kerr-N-418-171,2004-Kirkup-N-428-412, 2007-Reichenbach-N-488-1046,2010-Ni-PRE-82-066211, 2018-Avelino-EPL-121-48003, 2022-Avelino-PRE-105-024309,2020-Szolnoki-EPL-131-68001}. In its simplest version, this model describes a system formed by individuals of three different species $A$, $B$ and $C$, competing among themselves in a cyclic and non-hierarchical dominance scheme, as illustrated in Figure \ref{fig1}. In this scheme, $A$ dominates $B$, $B$ dominates $C$, and $C$ dominates $A$. This is the Standard RPS model, which we will refer to as the Std model throughout this work. It should be noted that the Std model exhibits symmetric interactions among species, which means that there are no privileged species in this case.

The RPS model has been extensively studied in the past few decades, since it captures a bunch of interesting features. In particular, in the spatial version of the RPS model, the individuals take place on the sites of a square lattice and the dynamics are usually implemented stochastically, via two possible dynamics: Lotka-Volterra and May-Leonard \cite{2008-Peltomaki-PRE-78-031906}. In this work, we have considered the May-Leonard dynamics, where individuals may move, reproduce, and compete. In addition, empty spaces are allowed on the lattice, which is fundamental for reproduction. The emergence of spatial structures and the maintenance of biodiversity are perhaps the most important features of this model.

{\color{black}Along the years, interesting work has been done, such as \cite{1995-Tanaka-PLA-207-53}, and more recently 
\cite{2001-Frean-PNAS-268-1323, 2019-Avelino-PRE-100-042209, 2022-Avelino-CSF-155-111738},
with the result that species that is least competitive may sometimes appear having the largest population. This result, known as the survival of the weakest \cite{2001-Frean-PNAS-268-1323}, has been further identified experimentally in investigations considering non-transitive asymmetric interactions among strains of E. coli \cite{2020-Liao-NC-11-6065}. The fact that the non-hierarchical approach represents an extreme simplification of most natural systems, in general opens to novel possibilities, among them the case where the species may not have the same strength. In order to accomplish this possibility and get closer to real-world systems, some models change the standard rules and take into account new features, including the fact that species may behave differently. In the previous works \cite{2019-Avelino-PRE-100-042209, 2022-Avelino-CSF-155-111738}, the authors considered examples of RPS-like models in the presence of the so-called {\it weak species}, i.e., species with weakened rates of reproduction and/or competition, finding the counter-intuitive behavior where the {\it weak species} may prevail over the others.}

In the same direction, in the present study we propose classes of models in which one or more species have higher reactive behavior in relation to the others. However, it is important to emphasize here that the mechanisms described in the present investigation are quite different from those considered in the related literature. In particular, a highly reactive species is the one formed by individuals that affect a larger region in its neighborhood, rather than having higher rates of reproduction and/or competition, and this is {\color{black}important distinction to be fully explored in this investigation, in the case of models controlled by three distinct species. To be more specific, in the present work we investigate two classes of RPS-like models, the so-called Highly Competitive (HC) and the Highly Reproductive (HR) models. The basic dynamics of such models are analyzed, and we also study how mobility may affect the probability of extinction in each of the two cases. This study may also be directly connected to other investigations of current interest, in particular, to the case investigated in Ref. \cite{2015-Szolnoki-EPL-110-38003}, where reentrant transitions appeared under defensive alliances in social dilemmas with informed strategies. Moreover, in the study of smart cooperators in a social dilemma situation described in \cite{2024-Hsuan-AMC-479-128864} which can unexpectedly thrive under high temptation, emphasizing the complexity of strategic interactions. This study suggests that the principles governing these interactions can be applied to other moral behaviors, such as truth-telling and honesty, providing valuable insights for future research in multi-agent systems and in the models to be investigated below.
Another possible line of investigation is related to real-life experiences, including key parameters characterizing cost of punishment, fines, and tax level, may be considered \cite{2024-Hsuan-CSF-178-114385}. This investigation determined how the spatiality of a population influences the competition of strategies when punishment is partly based on a uniform tax paid by all participants. The extension results in a more subtle system behavior in which different ways of coexistence can be observed, including dynamic pattern formation owing to cyclic dominance among competing strategies. This is also in direct connection with the present investigation, and can also be explored following the basic idea under the rules of the highly reactive environment which will be described in the present work.}

{\color{black} Another line of investigation concerns the complex interplay among ecological factors and the evolutionary dynamics of the species in different environments. Some issues have been recently explored in Refs. \cite{2024-Roy-PRSA-480-20240127,2025-Roy-PRSA-481-2307}, in particular, the case of multi-game framework that incorporates games that exhibit both transitive and cyclic dominance. By using analytical and numerical simulations, the authors explored evolutionary game dynamics under varying conditions, showing that communities can change from a transitive dominance to a cyclical structure, which significantly affects ecosystem community structures. An interesting result is that game-independent factors and intrinsic game parameters can drive a transition from transitive dominance to cyclic dominance, and we think this may be further explored including the basic idea under the rules of the highly reactive models which will be investigated in the present work.}

In order to accomplish some of the above possibilities, we have organized the work as follows. The next section is dedicated to the definition of the Std model, which will be our reference throughout the paper, as well as to the HC and HR models. In Sec. \ref{sec-meth} we introduce the basic methodology used to describe and study these models. The main results are presented in Sec. \ref{sec-results}, and we conclude the work with some comments and perspectives of future investigations in Sec. \ref{sec-conclusion}.

\section{Rock-Paper-Scissor Models with May-Leonard Dynamics}

The Std model, which we consider in this investigation, works according to the cyclic rules of the RPS game, following the May-Leonard dynamics. It takes place on a square lattice of size $L\times L$, with periodic boundary condition. Each site $i$ is occupied by an agent $\phi_i$ that can be  either an empty space or an individual of species $A$, $B$ or $C$, being $\phi_i=0,\; 1,\;2 \hbox{ or } 3$, respectively.

The dynamic follows the stochastic May-Leonard cyclic dominant framework, where the following steps are randomly taken at each iteration. 
\begin{enumerate}
    \item \label{step1} A site $i$ is selected.
    \item \label{step2} A site $j$ adjacent to (nearest neighboring) $i$ is selected.
    \item \label{step3} An action is selected and executed if it is possible to be executed. 
\end{enumerate}
One time step of such simulation, also known as a generation, is defined as $L^2$ repetitions of these steps. This is done in order to each site to be selected once on average. A whole simulation usually takes a sufficiently large number of generations (normally, thousands of generations), which may change from case to case.

May-Leonard dynamic contemplates three possible actions: movement, reproduction and competition, having each of the three a rate $pm$, $pr$ and $pc$, respectively, of being chosen. For the sake of normalization, it is required that $pm+pr+pc=1$. {\bf Movement} is the simplest action, in the sense that it can always be performed. It consists of simply swapping the positions of two adjacent agents $\phi_i$ and $\phi_j$. {\bf Reproduction}, on the other hand, occurs only when the adjacent site $j$ is empty. In that case a new agent of type $\phi_i$ is associated to site $j$. If reproduction is selected and the adjacent site $j$ is not empty, then nothing happens. {\bf Competition} only occurs when the adjacent agent $\phi_j$ is dominated by $\phi_i$. In that case, the adjacent site $j$ becomes an empty space. If the adjacent agent $\phi_j$ is not dominated by $\phi_i$, then nothing happens.

Throughout this work we have considered, without loss of generality, that reproduction and competition have the same rate, in other words, $pr=pc=\alpha$, being $\alpha$ a normalization factor. It is important to note that movement is a two-dimensional random walk process, and thus the rate $pm$ can be better parameterized by the mobility $M$ as $pm=\alpha 2 M L^2$ \cite{2007-Reichenbach-N-488-1046, 2017-Bazeia-EPL-119-58003}. Using the normalization condition $pm+pr+pc=1$, one can obtain $\alpha$ and the rates are then given by 
\begin{eqnarray}
    pm &=& \frac{ML^2}{1+ML^2}\\
    pr = pc &=& \frac{1/2}{1+ML^2}
\end{eqnarray}
where the mobility $M$ represents a parameter ranging from zero to infinity. 

In this work, we propose two different families of RPS models, which can be seen as extensions of the above Std model, considering the presence of reactive species, which can be either highly competitive or highly reproductive.

\subsection{Highly Competitive Models}
\begin{figure}[!htp]
	\centering
	\includegraphics[scale=1]{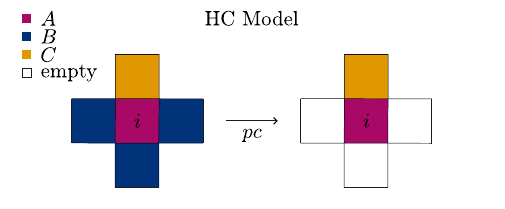}
	\caption{Illustration of an example where species $A$ is highly competitive. In this case, whenever an individual of species $A$ located at a site $i$ predates, it dominates not only one but all individuals of specie $B$ neighboring site $i$. We refer to the model where only species $A$ is highly competitive as HC-$A$ model, meaning that species $B$ and $C$ act just as in the Std model.}
	\label{fig2}
\end{figure}

The family of HC models takes into account the presence of highly competitive species. To be more specific, highly competitive species are those competing not only at the selected adjacent site, but at all the sites neighboring $i$. In Figure \ref{fig2}, an illustration is shown in which species $A$ is highly competitive. We refer to the model where only species $A$ is considered highly competitive as HC-$A$. In the HC-$A$ model, the other species $B$ and $C$ follow the standard competition rules. Likewise, we refer to the model in which species $A$ and $B$ are highly competitive as HC-$AB$ and HC-$ABC$ to the model in which all species are highly competitive. It is important to notice here that in HC models, only HC-$ABC$ exhibits interaction symmetry among the species, in the sense that there is no privileged species.

\subsection{Highly Reproductive Models}
\begin{figure}[!htp]
    \centering
    \includegraphics[scale=1]{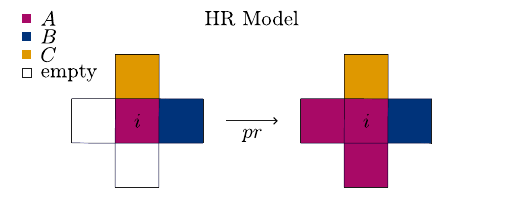}
    \caption{Illustration of an example where species $A$ is highly reproductive. In this case, whenever an individual of species $A$ located at a site $i$, reproduces, it makes a copy of itself at all empty sites neighboring site $i$. We refer to the model where only species $A$ is highly reproductive as HR-$A$ model, meaning that species $B$ and $C$ act just as in the Std model.}
    \label{fig3}
\end{figure}

Similarly to HC models, the family of HR models takes into account the presence of highly reproductive species. In other words, species that reproduce not only in the selected adjacent site, but in all empty spaces adjacent to the site $i$. In Figure \ref{fig3}, an illustration can be seen in which species $A$ are highly reproductive. The notation used for HR models follows the same pattern as used for HC models, i.e., we refer to the model where only species $A$ is highly reproductive as HR-$A$ model, and in this case, the other species follows the standard reproduction rules. The case where species $A$ and $B$ are highly reproductive is referred to as the HR-$AB$ model and, finally, HR-$ABC$ refers to the model where all species are highly reproductive, which again is the only HR model exhibiting interaction symmetry among species.

\section{Methodology}
\label{sec-meth}
The methodology used to analyze the two families of models proposed in this work follows the same path used to analyze the Std model, via stochastic simulations. {\color{black}The main code used for all simulations displayed in this work, as well as the codes to analyze the results, are all available in Ref. \cite{code}.} The first investigation is conducted by simulating these models on a square lattice sized $L=500$ and $M=4\times 10^{-6}$. Such a choice of mobility is made in order to obtain $pm=0.5$ and $pr=pc=0.25$. As will become clear soon, starting from a random initial condition, all models presented in this work are stable after about $1000$ generations, thus $6000$ generations is a long enough simulation time.

An initial and useful qualitative analyzes can be obtained by looking at the snapshots of the spatial structures after a long time simulation. Important aspects such as the emergency of spatial patterns can be identified at this stage. Next, the analysis can be made more quantitative by looking at the time evolution of the density of species $\rho_\mu(t)$, defined as the number of agents of species $\mu$ at time $t$ divided by $L^2$, which is $\mu= A, B, C \hbox{ or } E$, where $\mu=E$ stands for empty space.

Usually, the species density oscillates over time around an average value with a characteristic frequency and a characteristic amplitude, where the former represents how fast the dominance among species interchanges over time. Both the frequency and amplitude of oscillations of a given species $\mu$ can be quantified by the power spectrum $\langle \vert \rho_\mu(f) \vert^2 \rangle$, where $\rho_\mu(f)$ represents the discrete Fourier transform of $\rho_\mu(t)$ at a frequency $f$, defined as
\begin{equation}
    \label{eq-powerespc}
    \rho_\mu(f) = \frac{1}{N_G}\sum_{t=0}^{N_G-1} \rho_\mu(t) e^{-i 2\pi f t }\;,
\end{equation}
being $f$ the frequency and $N_G$ the number of generations. In this work we computed the power spectrum of the species averaged over 1000 simulations for each model.

Another important quantitative aspect of such models is the average size of the spatial structure, also known as the characteristic length $l$. These lengths can be quantified by the autocorrelation function $C(r)$ that measures the correlation between agents $r$ units of length apart. The precise definition of the autocorrelation function used in this work is given below.  

Let $\mathcal{L}$ be a square lattice. The distance between two sites $i$ and $j$ is represented by $d(i,j)$ and is by definition the least number of edges connecting the two sites. For a given site $i$ let $N_r(i)$ be the set of all sites whose distance from $i$ is equal to $r \Delta r$, in other words
\begin{equation}
    N_r(i) = \left\{ j\in \mathcal{L} : d(i,j)= r \Delta r \right\}
\end{equation}
being $r=0,1,2,3,\cdots, L$ and $\Delta r$ the grid spacing, which here we will set as $\Delta r = 1$ without lost in generality.

The autocorrelation coefficient between sites $r$ ($\neq 0$) units of distance apart is defined by
\begin{equation}
    \label{eq-sss1}
    f_r =\frac{1/2}{r+1} \sum_{i\in \mathcal{L}} \phi_i \left( \sum_{j\in N_r(i)}  \phi_j    \right)\;,
\end{equation}
where the factor $1/2$ is due to the fact that every pair of sites are counted twice for $r\neq 0$. For $r=0$, on the other  hand, the autocorrelation coefficient is simply
\begin{equation}
    \label{eq-sss2}
    f_0 = \sum_{i\in \mathcal{L}} \phi_i^2\;.
\end{equation}
Finally, the autocorrelation function is given by the ratio 
\begin{equation}
    \label{eq-autocor}
    C(r) = \frac{f_r}{f_0}\;,
\end{equation}
Note that $C(0)=1$.

One of the key features of RPS like models is the existence of a stable phase supporting the co-existence of all species. However, all these models may eventually end up in an extinction phase with a certain probability, called the probability of extinction. It is known from the literature that in the Std model this probability vanishes for sufficient small values of mobility, meaning that small mobility leads to better chances of surviving, while higher mobility jeopardizes biodiversity \cite{2007-Reichenbach-N-488-1046}. In this work, we also analyze how this probability changes in highly reactive models compared to mobility. This is done by simulating the models $1000$ times and counting the number of simulations that ended up in extinction. 

\section{Results}
\label{sec-results}

\subsection{Highly Competitive Models}

\begin{figure}[!htp]
    \centering
    \includegraphics{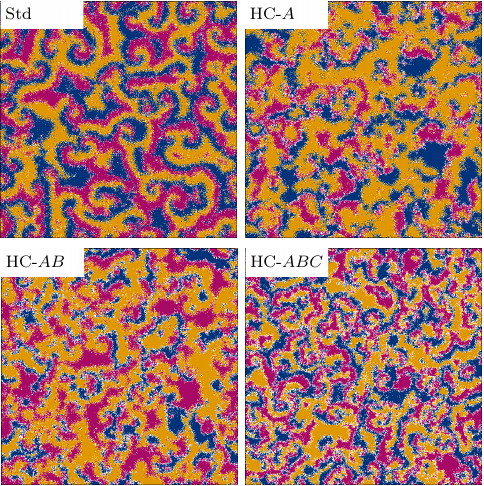}
    \caption{{\color{black}{Snapshots of both Std and HC models after $6000$ generations, highlighting the emergence of spatial patterns. For these simulations we have set }} $L=500$  and $M=4\times 10^{-6}$, meaning that $pm= 0.5$ and $pc = pr = 0.25$. The colors of species $A$, $B$ and $C$ follows the same color palette used in Figure \ref{fig1}, while empty spaces are represented in white color. (a) Std model - the spiral patterns are quite evident in this case, also the system exhibits a certain homogeneity, in the sense that the  density of all three species seems to be more or less the same, leading to a non-hierarchical behavior. (b) HC-$A$ model - in this case it is possible to visually estimate that species $C$ prevail over the other species. (c) HC-$AB$ - just as before, in this case it is possible to see that species $B$ is at disadvantage compared to the others. Both case, (b) and (c), are example of hierarchical models. (d) HC-$ABC$ - in this case the snapshots suggest again an homogeneity of the system, leading to a non-hierarchical behavior.
    }
    \label{fig4}
\end{figure}

Let us first look at the HC models by simulating the models with fixed parameters $L=500$ and $M=4\times 10^{-6}$. The resulting snapshots of these simulations after $6000$ generations are shown in Figure \ref{fig4}. The first thing to be noticed from the snapshots is that all HC models support biodiversity for a long period of time for the chosen parameters. The snapshots allow us to visualize that spiral patterns are quite evident in the symmetric models, as is the case for Std and HC-$ABC$ models. The emergency of spiral patterns in the Std model is known from the literature, but here one can notice that the HC-$ABC$ model also exhibits these patterns, though in this case they are not as evident as in the Std model. In asymmetric models, on the other hand, such as HC-$A$ and HC-$AB$, one can notice the absence of spatial patterns, which means that the spatial patterns are destroyed in the presence of asymmetric interactions.

\begin{figure}[!htp]
    \centering
    \includegraphics{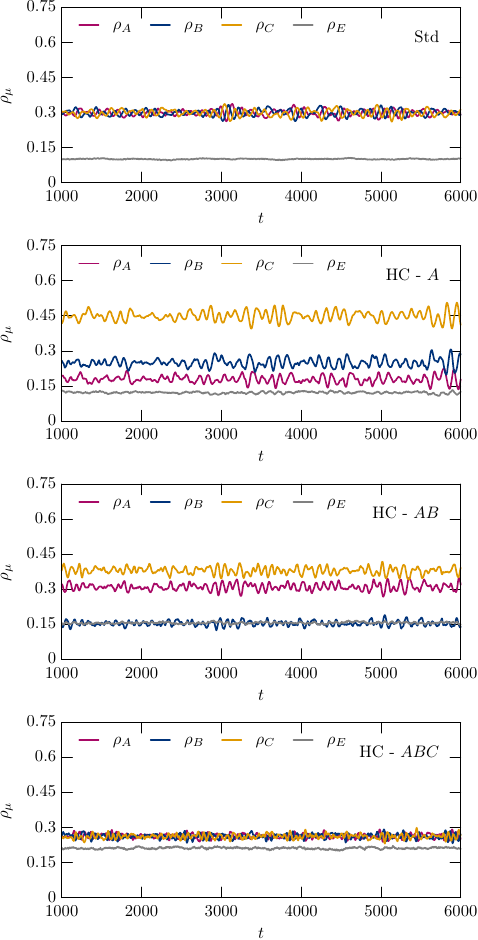}
    \caption{
        Time evolution of species densities displayed for over $6000$ generations for both Std and HC models. {\color{black}{The graphs help visualize the disadvantage faced by the highly competitive species in non-symmetrical models. Also, symmetrical models exhibits non-hierarchical dominance.}} Here the lattice size is set as $L=500$ and $M=4\times 10^{-6}$, meaning that $pm= 0.5$ and $pc = pr = 0.25$. The first $1000$ generations include the unstable phases before the system reaches the equilibrium, so they were left out of the analysis since we are only interested at the equilibrium phase of such models.}
    \label{fig5}
\end{figure}

Another behavior to be noticed from the snapshots in Figure \ref{fig4} is that the species density of the symmetric models are the same on average, which means that no species dominates the system for too long. In contrast, snapshots suggest that species $C$ dominates over the others in both asymmetric models, HC-$A$ and HC-$AB$. These results can be made more quantitative by looking at the time evolution of the species density, shown in Figure \ref{fig5}.

In fact, it can be observed from Figure \ref{fig5} that both symmetric models exhibits the same mean value for all species density. In the HC-$A$ model, for example, where species $A$ is highly competitive, the second dominant species is species $B$, being $A$ the least dominant of all species. In HC-$AB$ model the least two dominant species are interchanged in relation to HC-$A$ model. These results are in agreement with Ref. \cite{2019-Avelino-PRE-100-042209}, where the authors have shown that in the presence of one strong species (as in HC-$A$) the dominance order is $C$, then $B$ and then $A$, while in the presence of two strong species (as in HC-$AB$) the dominance order is $C$, then $A$ and then $B$, just as happens in the HC models.

The presence of a highly competitive species, as $A$ in HC-$A$ model, causes the density of $B$ species to decrease, thus increasing the density of $C$ species. The high density of $C$ will then make the density of the species $A$ decrease as well. However, even for low density values for species $A$, the density of $B$ species will remain low, due to the highly competitive behavior of species $A$. In summary, species $C$ will always have an advantage over species $A$ and $B$ in this model, just as observed in Figure \ref{fig5}.

It is also important to note that the mean value for the density of empty spaces increases with the number of highly competitive species, since the number of competition also increases. In this sense, the HC-$ABC$ model has on average more empty spaces than the Std model. Lastly, Figure \ref{fig5} also allows us to see that the average species density of each species deviates more evidently from the mean value in the asymmetric models. For instance, in the HC-$A$ model this deviation appears to be the highest, when compared to the other models displayed in Figure \ref{fig5}, making this HC-$A$ more likely to reach extinction, as we will see in detail later. 

\subsection{Highly Reproductive Models}

\begin{figure}[!htp]
    \centering
    \includegraphics{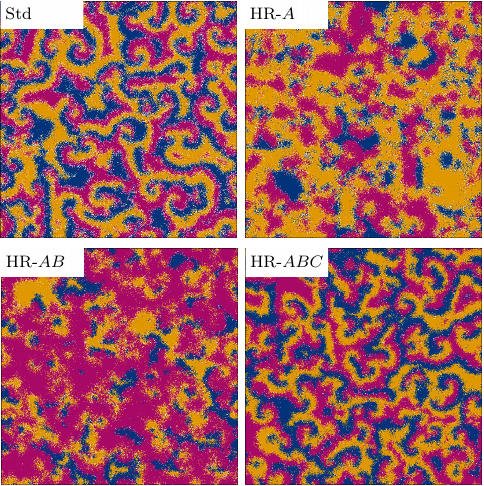}
    \caption{{\color{black}{Snapshots of both Std and HR models after $6000$ generations, highlighting the emergence of spatial patterns. For these simulations we have set }} $L=500$  and $M=4\times 10^{-6}$, meaning that $pm= 0.5$ and $pc = pr = 0.25$. The colors of species $A$, $B$ and $C$ follows the same color palette used in Figure \ref{fig1}, while empty spaces are represented in white color. (a) Std model - the spiral patterns are quite evident in this case, also the system exhibits a certain homogeneity, in the sense that the  densities of the three species seem to be more or less the same, leading to a non-hierarchical behavior. (b) HR-$A$ model - in this case it is possible to visually estimate that species $C$ prevail over the other species. (c) HR-$AB$ - just as before, in this case it is possible to see that species $A$ prevails over the others. Both (b) and (c) models are examples of hierarchical models. (d) HR-$ABC$ - in this case the snapshot suggests again a certain homogeneity present in the system, leading to a non-hierarchical behavior.}
    \label{fig6}
\end{figure}
\begin{figure}[!htp]
    \centering
    \includegraphics{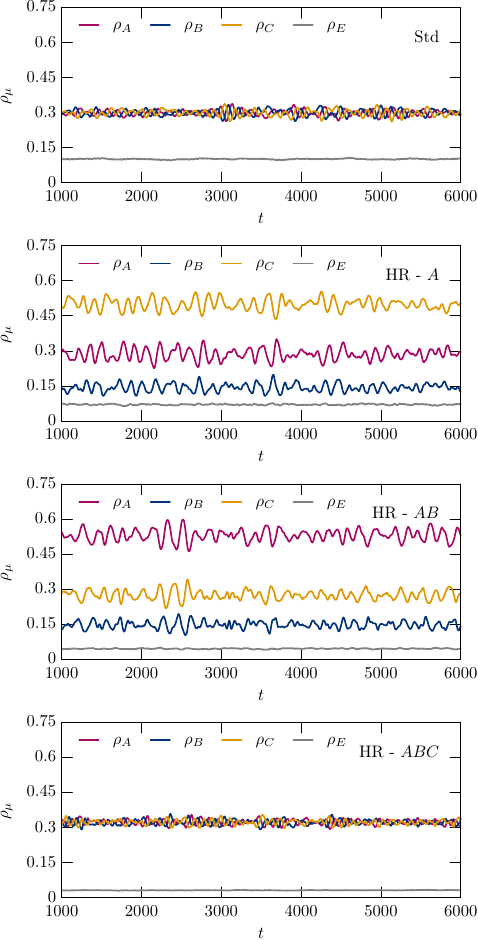}
    \caption{
        Time evolution of species densities displayed for over $6000$ generations for the Std and the HR models. {\color{black}{The graphs help visualize the advantage/disadvantage faced by the highly reproductive species in non-symmetrical models. Also, symmetrical models exhibits non-hierarchical dominance.}} Here the lattice size is set as $L=500$ and $M=4\times 10^{-6}$, meaning that $pm= 0.5$ and $pc = pr = 0.25$. The first $1000$ generations include the unstable phases, before the system reaches the equilibrium, so they were left out of the analysis since we are only interested at the equilibrium phase of such models.}
    \label{fig7}
\end{figure}

When investigating the HR models one may find similarities and distinctions if compared with the previous models. In order to highlight the distinctions, we have simulated HR models with the same parameters used for HC models, and the resulting snapshots (after 6000 generations) are the ones shown in Figure \ref{fig6}. In addition, the time evolution of species density for these HR models is shown in Figure \ref{fig7}. 

The snapshots in Figure \ref{fig6} clearly show that all HR models support biodiversity for a long period of time, just as the HC models. In addition, spiral patterns are only evident in the HR-$ABC$ model, which is the only symmetric HR model. Moreover, it is important to highlight that the spiral patterns are more evident in HR-$ABC$ than in HC-$ABC$.

The time evolution of the species densities shown in Figure \ref{fig7} shows that in symmetric models the mean value of the species density is the same value for all species, as expected. The asymmetric models, on the other hand, exhibit different mean values for species density. These models again are in agreement with the literature \cite{2019-Avelino-PRE-100-042209}, where the authors have shown that in the presence of strong species with respect to reproduction (as it is the case for HR-$A$ and HR-$AB$), the densities exhibit the same dominance ordering of the HR models.

It is important to note that the density of empty spaces decreases as the number of highly reproductive species increases, as expected, since the number of reproductions increases in these models. 

\subsection{Power Spectrum}

The density functions $\rho_\mu(t)$ shown in Figures \ref{fig5} and \ref{fig7} oscillate around a mean for all models and species, which is typical of RPS-like models. However, the characteristic frequency and amplitudes of such functions are not clear from these pictures. For this reason, let us now look at the power spectrum of each species. In order to use Eq. \eqref{eq-powerespc}, one has to specify the number of generations used in the computations, which in our case is $N_G=5000$, since simulations here are considered from $1000$ to $6000$ generations.

\begin{figure}[!htp]
    \centering
    \includegraphics{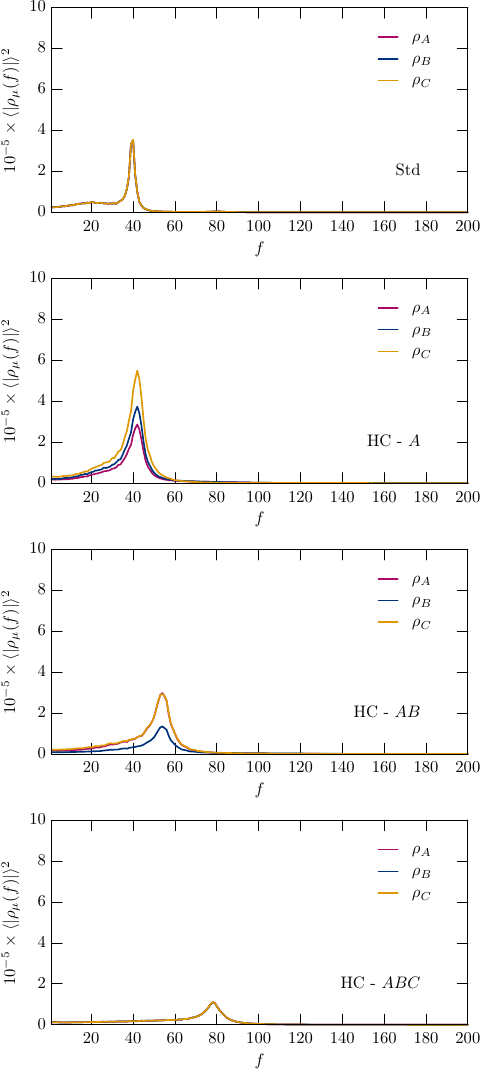}
    \caption{
        Power spectrum of each species in both Std and HC models. {\color{black}{The graphs display the characteristic frequency of each species. One notices that the characteristic frequency increases with the number of highly competitive species. The differences in peak heights are due to variations in the oscillation amplitudes of the species.}} Here we have set $L=500$ and $M=4\times 10^{-6}$, meaning that $pm= 0.5$ and $pc = pr = 0.25$ for all analyzed models. Each power spectrum displayed in this picture was obtained by the time evolution of the species density using the prescription outlined in section \ref{sec-meth}. The species densities, in turn, were averaged over $1000$ simulations and $6000$ generations, where the first $1000$ generations were left out of the analysis since they represent the non-equilibrium phase in these models.
    }
    \label{fig8}
\end{figure}

In Figure \ref{fig8}, the power spectrum for the Std and the HC models is shown.
The first feature to be noticed from Figure \ref{fig8} is that in HC models all species have the same characteristic frequency, which suggests that it may be a model property. The characteristic frequency for the Std model, for example, is close to $40$, meaning the species dominance interchanges about $40$ times for each $5000$ generations. Also, it can be noticed from Figure \ref{fig8} that the characteristic frequency increases with the number of highly competitive species, in other words, HC-$ABC$ is the one with the higher characteristic frequency, meaning that the dominance of species interchanges faster than any other HC model, as well as the Std model, and this seems to be mostly due to the fact that competition happens more often in these models. 

The highness of a peak in the power spectrum is associated to the amplitude of oscillation in the corresponding frequency. Again, as expected, all species exhibit the same power spectrum in the symmetric models, though the peaks in HC-$ABC$ are not as high as in the Std model, meaning the amplitudes associated with the characteristic frequency are lower for the HC-$ABC$ model than it is for the Std model, which is in agreement with Figure \ref{fig5}. In asymmetric models, on the other hand, the peaks do not have all the same height, meaning that the species oscillate with distinct amplitudes. In the HC-$A$ model, for example, species $C$ is the one with the highest amplitude, with species $A$ being the one with the lowest amplitude. Again, this is in agreement with Figure \ref{fig5}, though it may be a bit hard to visualize such a feature from that picture. In the HC-$AB$ model, species $A$ and $C$ are the ones with higher amplitudes, while species $B$ is the one with the lowest oscillation amplitude. As said before, the asymmetric HC models are those where the average species density deviates the most from the mean value, and this deviation decreases as the number of highly competitive species increases.

\begin{figure}[!htp]
    \centering
    \includegraphics{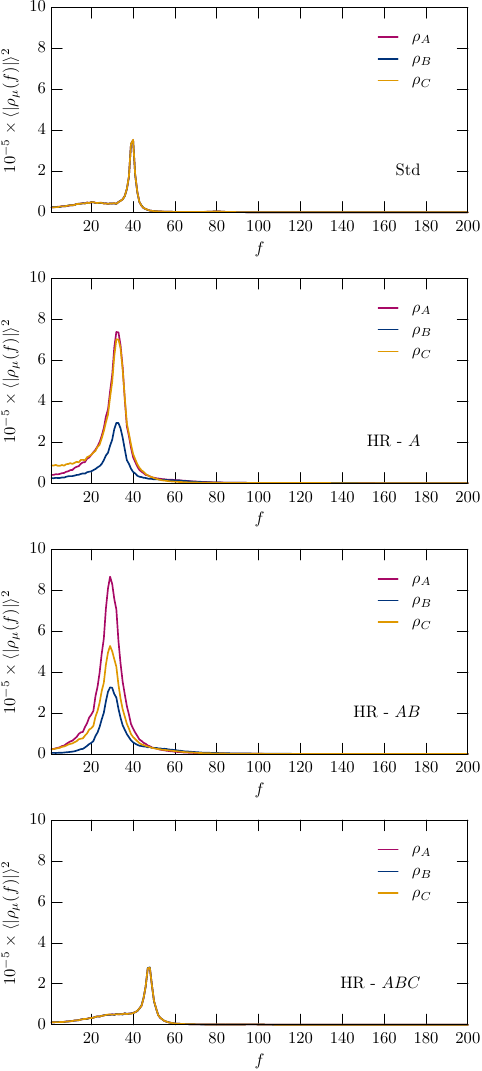}
    \caption{
        Power spectrum of each species in both Std and HR models. {\color{black}{The graphs display the characteristic frequency of each species. The differences in peak heights are due to variations in the oscillation amplitudes of the species.}} Here we have set $L=500$ and $M=4\times 10^{-6}$, meaning $pm= 0.5$ and $pc = pr = 0.25$, for all analyzed models. Each power spectrum displayed in this picture was obtained by the time evolution of the species density using the prescription outlined in section \ref{sec-meth}. The species densities, in turn, were averaged over $1000$ simulations and $6000$ generations, where  the first $1000$ generations were left out of the analysis since they represent the non-equilibrium phase in these models.
    }
    \label{fig9}
\end{figure}

The power spectrum of the HR model is shown in Figure \ref{fig9}. Again, it is clear from the picture that the characteristic frequency is a model property, meaning that all species have the same characteristic frequency. Also, the power spectrum are identical for all the species in symmetric models. 

However, unlike the HC models, the characteristic frequency in HR models does not seem to change much. The asymmetric models seem to have almost the same characteristic frequency, which is, by the way, a bit lower than the characteristic frequency of the Std model, while the symmetric HR-$ABC$ model is the one with the highest characteristic frequency, but still not as high as in the HC-$ABC$ model.

It is possible to see by looking at the power spectrum that the peaks in the HR-$ABC$ model are pretty much as high as in the Std model, which means there is not much difference in their amplitudes of oscillation. In the asymmetric models, on the other hand, the differences are even more evident than in HC models, which can be compared to the amplitudes of oscillation in Figure \ref{fig7}.

\subsection{Autocorrelation Function}

The average size of spatial structures can be quantified by the characteristic length $l$, which plays an important role in these models. In this sense, the autocorrelation function $C(r)$, in Eq. \eqref{eq-autocor}, allows us to define a $l$ such that the autocorrelation function is $20\%$, in other words, such that $C(l)=0.2$. A similar approach was used before in Ref. \cite{2018-Avelino-PRE-97-032415}. The characteristic length can be used in order to estimate the probability of extinction, in the sense that the probability of extinction tends to increase as the characteristic length gets closer to the grid size. This relation is more evident in the symmetric models, since in these cases the characteristic length of each species is more or less the same.

\begin{figure}[!htp]
    \centering
    \includegraphics{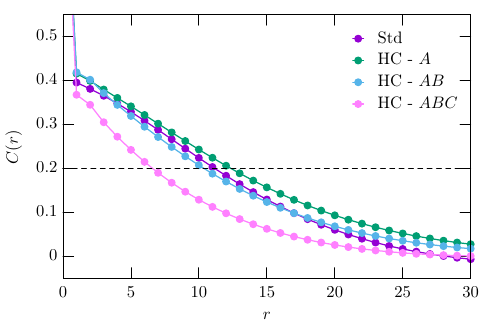}
    \caption{
        Autocorrelation function of both Std and HC models. Here we have performed $1000$ simulations evolved for over $6000$ generations, where the first $1000$ generations were left out of the analysis in order to skip the transient part of the time evolution. For all simulations we have set $L=500$ and $M=4\times 10^{-6}$, meaning $pm= 0.5$ and $pc = pr = 0.25$. Also, each autocorrelation function is averaged over all simulation performed. The horizontal dashed line represents our choice for the autocorrelation function in order to define the characteristic length $l$, in other words, as the solution for the equation $C(l)=0.2$. {\color{black}{As it can be noticed the characteristic length is smaller when all species are considered highly competitive.}} 
    }
    \label{fig10}
\end{figure}

In Figure \ref{fig10}, the autocorrelation function of the HC models is shown. It is not clear from the picture which asymmetric model has a longer/smaller characteristic function compared to Std model. This is due to the high heterogeneity these models exhibit, meaning that the different species exhibit different structure sizes. However, as can be observed, the characteristic length in the HC-$ABC$ model is shorter than in the Std model. In this sense, it is plausible to expect that the HC-$ABC$ supports biodiversity for higher mobility than the Std model does.

\begin{figure}[!htp]
    \centering
    \includegraphics{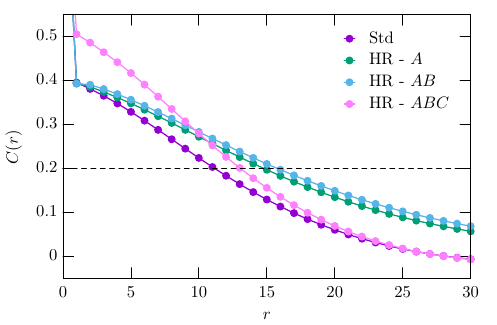}
    \caption{
        Autocorrelation function of both Std and HR models. Here we have performed $1000$ simulations evolved for over $6000$ generations, where the first $1000$ generations were left out of the analysis in order to skip the transient part of the time evolution. For all simulations we have set $L=500$ and $M=4\times 10^{-6}$, meaning $pm= 0.5$ and $pc = pr = 0.25$. Also, each autocorrelation function is averaged over all simulation performed. The horizontal dashed line represents our choice for the autocorrelation function in order to define the characteristic length $l$, in other words, as the solution for the equation $C(l)=0.2$. {\color{black}{As it can be noticed the characteristic length is smaller in the absence of highly reproductive species.}}
    }
    \label{fig11}
\end{figure}

Another somewhat distinct behavior is observed in HR models. Figure \ref{fig11} shows the autocorrelation function for these models. As can be noticed, again, the asymmetric models cannot be properly differentiated from the Std model when it comes to the characteristic length due to the high heterogeneity of species. Although the HR-$ABC$ model and Std model exhibit a similar spiral pattern, HR-$ABC$ model is the one exhibiting a slightly longer characteristic length. This is basically due to the density of empty space, since the Std Model has a higher density of empty space which leads to a lower value for the autocorrelation function for small $r$ (smaller than $20$). On the other hand, for $r>20$ both models have the same autocorrelation function. In this case, the average spiral sizes are the same for both models as well as the probability of extinction, as shown in figure \ref{fig13}.  

\subsection{Probability of Extinction}

\begin{figure}[!htp]
    \centering
    \includegraphics{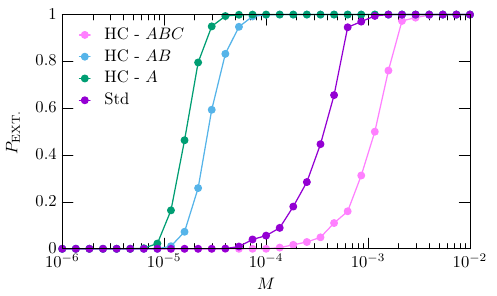}
    \caption{    
        Probability of extinction as a function of mobility for both Std and HC models. Here we have performed $1000$ simulations evolved for over $40000$ generations for each value of $M$, varying from $10^{-6}$ to $10^{-2}$ with the interval uniformly divided in 30 parts on a logarithmic scale. The first $1000$ generations were left out of the analysis in order to skip the transient part of the time evolution. For all simulations we have set $L=200$. The probability of extinction is then obtained by counting the number of simulations led compared with the number of simulations performed. {\color{black}{As it can be noticed the asymetrical models are the ones with higher probability of extinction. Additionally, the presence of highly competitive species in the HC-$ABC$ model makes it more robust to extinction compared to the Std Model.}}
    }
    \label{fig12}
\end{figure}

\begin{figure}[!htp]
    \centering
    \includegraphics{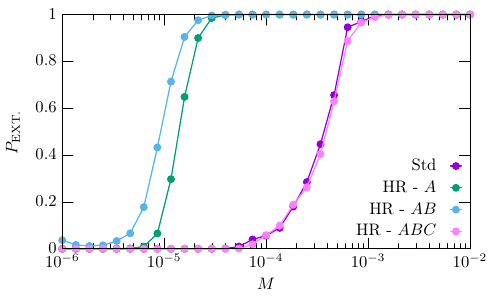}
    \caption{
        Probability of extinction as a function of mobility for both Std and HR models. Here we have performed $1000$ simulations evolved for over $40000$ generations for each value of $M$, varying from $10^{-6}$ to $10^{-2}$ with the interval uniformly divided in 30 parts on a logarithmic scale. The first $1000$ generations were left out of the analysis in order to skip the transient part of the time evolution. For all simulations we have set $L=200$. The probability of extinction is then obtained by counting the number of simulations led compared with the number of simulations performed. {\color{black}{As it can be noticed the asymetrical models are the ones with higher probability of extinction. Additionally, the presence of highly competitive species in the HR-$ABC$ model makes it as robust to extinction as the Std Model.}}
    }
    \label{fig13}
\end{figure}

 The assumptions made above for the systems supporting biodiversity as a function of mobility need to be explored in more detail. To implement this possibility, here we have simulated both the HC and HR models for different values of mobility to compute the corresponding probabilities of extinction. For each value of mobility, we have simulated the models for more than $1000$ simulations on a square lattice of size $L=200$. The results are shown in Figures \ref{fig12} and \ref{fig13} for the HC and HR models, respectively.

In Figure \ref{fig12}, it is shown that the HC-$ABC$ model supports biodiversity for the highest mobility values, which means that the symmetric model in the presence of highly competitive species has better chances of survival. Also, as expected, the asymmetric models are the ones with lower possibility to support biodiversity, meaning that non-hierarchical interactions lead the systems to better chances of survival.

A similar behavior happens for the HR models. Again, asymmetric models are those with lower capacity to survive, i.e., they only support biodiversity for small values of mobility. The symmetric model HR-$ABC$, on the other hand, is very similar to Std model when it comes to the capacity to survive.

\section{Ending comments}
\label{sec-conclusion}

In this work, we have investigated several aspects of two classes of highly reactive models, the highly competitive (HC) and the highly reproductive (HR) models, in comparison with the standard model (Std model) based on the rock-paper-scissors rules under the May-Leonard dynamics. We have considered the simplest system, composed of three distinct species $A$, $B$, and $C$ that evolves in a spatial lattice of size $500\times 500$, using periodic boundary conditions. We have studied the Std model, the HC-$A$ and HR-$A$, the HC-$AB$ and HR-$AB$, and the HC-$ABC$ and HR-$ABC$ models. The presence of $A$, or $AB,$ or $ABC$ in these models means that the species $A$, or $A+B$, or $A+B+C$ have enlarged competitive or reproductive rules in the HC or HR models, respectively. In Figures \ref{fig4} and \ref{fig6} we have displayed snapshots of time evolution after $6000$ generations for the Std model and for the HC-$A$, HC-$AB$, HC-$ABC$, and for the Std model and the HR-$A$, HR-$AB$, HR-$ABC$ models, respectively. These models have a distinct evolution, so we have also investigated the evolution of species density in Figures \ref{fig5} and \ref{fig7}, the power spectrum in Figures \ref{fig8} and \ref{fig9}, and the autocorrelation function in Figures \ref{fig10} and \ref{fig11}. Finally, in Figures \ref{fig12} and \ref{fig13}, we have added results for the probability of extinction as a function of mobility, in this last case in a lattice of $200\times200$.

Among the several results, we want to highlight the fact that, in relation to the maintenance of biodiversity as a function of mobility, the presence of asymmetry tends to weaken biodiversity. Moreover, in the symmetric models, that is, in the Std model and in the HC-$ABC$ and HR-$ABC$ models, when the highly reactive model is controlled by reproduction, one notices the absence of modification in the probability of extinction, but the behavior changes when the reactive rule is competition. In this case, the most robust model is the HC-$ABC$. This means that the symmetric increasing of competition tends to fortify biodiversity. This seems to be a result of current interest in the study of the relationships between living organisms in nature. 

There are distinct possibilities of investigations related to the results of the present work, one of them being the enlargement of the system, adding more species and new rules of competition. Concrete examples could be to consider the cases of four and five species, in which we have to add interactions among the second neighbor, present in the two cases. Since we are adding new rules of competition, it is possible that new results will emerge, in direct connection with the behavior found in the present work. {\color{black} We can also consider using ideas similar to the case of the highly reactive rules utilized in the present work to explore other systems, in particular, the models of social dilemmas investigated in \cite{2015-Szolnoki-EPL-110-38003,2024-Hsuan-AMC-479-128864,2024-Hsuan-CSF-178-114385}, and the study of ecological factors and evolutionary dynamics of species under the lines of Refs. \cite{2024-Roy-PRSA-480-20240127,2025-Roy-PRSA-481-2307}}. We are now studying some possibilities, hoping to return to this issue in the near future.

\begin{center}
{\bf Acknowledgments}
\end{center}

This work is supported by Conselho Nacional de Desenvolvimento Científico e Tecnológico (CNPq, Grants 303469/2019-6 (DB), 402830/2023-7 (DB), 309835/2022-4 (BFO)), Fundação de Apoio a Pesquisa do Estado da Paraíba (FAPESQ-PB, Grant 0015/2019) and INCT-FCx (CNPq/FAPESP).

%
\end{document}